# Two-Potential Formalism for Numerical Solution of the Maxwell Equations

S. I. Trashkeev,* A. N. Kudryavtsev**

*Institute of Laser Physics, Siberian Branch, Russian Academy of Sciences (Novosibirsk)
**Khristianovich Institute of Theoretical and Applied Mechanics, Siberian Branch, Russian Academy of Sciences (Novosibirsk)
e-mails: sitrskv@mail.ru; alex@itam.nsc.ru

Abstract

A new formulation of the Maxwell equations based on two vector and two scalar potentials is proposed. The use of these potentials allows the electromagnetic field equations to be written in the form of a hyperbolic system. In contrast to the original Maxwell equations, this system contains only evolutionary equations and does not include equations having the character of differential constraints. This fact makes the new equations especially convenient for numerical simulations of electromagnetic processes; in particular, they can be solved by modern powerful shock-capturing methods based on approximation of spatial derivatives by upwind differences. The electromagnetic field both in vacuum and in an inhomogeneous material medium is considered. Examples of modeling the propagation of electromagnetic waves by means of solving the formulated system of equations with the use of modern high-order schemes are given.

**Key words:** computational electrodynamics, two-potential formalism, numerical solution of hyperbolic systems of equations, shock-capturing schemes.

## 1. INTRODUCTION

The Maxwell equations form the basis of the classical electrodynamics, which describes an extremely wide range of physical phenomena observed in nature and used in various engineering devices. Computer modeling of propagation of electromagnetic waves and their interaction with matter, based on the numerical solution of the Maxwell equations, has become one of the most important tools of research in many fields of science. It is sufficient to tell that three editions of the book of *A. Taflove and S.G. Hagness* [1] dealing with only one (though the most popular one) numerical method of solving these equations were published during the last decade, and the last edition has more than a thousand pages. Particularly rapid development of computational electrodynamics has been recently observed owing to its applications in vigorously progressing important fields such as nonlinear optics, laser physics, nanophotonics, and creation of new materials (photon crystals, "soft matter," including liquid crystals and various "smart" media).

During many years that passed since *J.C. Maxwell* derived equations that are known now under his name, versatile formulations were proposed: from the original form in the components of the electric and magnetic fields [2] to modern approaches involving the differential forms, the theory of spinors, and the Clifford algebra [3]. Being equivalent from the physical viewpoint, different formulations are often inequivalent from the viewpoint of convenience of their use in practice, in particular, in numerical simulations of electromagnetic processes. The choice of an appropriate formulation can appreciably simplify the algorithm used for numerical modeling and significantly increase the accuracy of the results obtained. Therefore, the search for the "best" formulation will be undoubtedly continued in the future. In this paper, we propose a new formulation of electromagnetism equations, which involves two vector potentials and two scalar potentials. In our opinion, this formulation is elegant, possesses some desirable mathematical properties, and offers important advantages for constructing high-accuracy numerical algorithms on its basis. In particular, this formulation allows well-developed (mainly, in computational fluid dynamics) methods of solving hyperbolic systems of equations to be used to the best possible extent in solving electrodynamic problems.

The outline of this paper is as follows. The two-potential formalism for electromagnetism equations in vacuum is described in Section 2. In Section 3, it is generalized to the case of an isotropic material medium. The mathematical properties of the derived equations are considered in Section 4. Examples of using the new approach for numerical simulation of electrodynamic processes are further given in Section 5. Finally, some conclusions are drawn in Section 6.

2. TWO-POTENTIAL FORMALISM FOR THE MAXWELL EQUATIONS IN VACUUM

The Maxwell equations in an empty space can be written as follows [4]:

$$\frac{\partial \mathbf{E}}{\partial t} - \nabla \times \mathbf{B} = -\mathbf{J}_e, \tag{2.1}$$

$$\frac{\partial \mathbf{B}}{\partial t} + \nabla \times \mathbf{E} = -\mathbf{J}_m, \tag{2.2}$$

$$\nabla \cdot \mathbf{E} = \rho_e, \tag{2.3}$$

$$\nabla \cdot \mathbf{B} = \rho_m. \tag{2.4}$$

Here, in addition to the electric charges $\rho_e$ and currents $\mathbf{J}_e$, we also retain the hypothetical magnetic charges $\rho_m$ and magnetic currents $\mathbf{J}_m$. Though magnetic charges and currents have never been observed in experiments, they arise almost in all modern fundamental theories of the matter [5]; therefore, it is rather probable that they do exist in nature. In writing Eqs. (2.1)-(2.4), the measurement units are chosen in such a manner that the speed of light, as well as the vacuum permittivity and permeability are equal to unity.

It follows from Eq. (2.4) that the magnetic field in the absence of magnetic charges is solenoidal, and it can be expressed via the magnetic vector potential as

$$\mathbf{B} = \nabla \times \mathbf{A}. \tag{2.5}$$

Substituting Eq. (2.5) into Eq. (2.2), we find that the curl of the vector $\partial \mathbf{A} / \partial t + \mathbf{E}$ vanishes in the absence of magnetic currents; therefore, this vector can be expressed via the gradient of the scalar electric potential $\varphi$:

$$\mathbf{E} = -\nabla \varphi - \frac{\partial \mathbf{A}}{\partial t}. \tag{2.6}$$

These are usual expressions for fields in terms of the vector and scalar potentials. On the other hand, if there are no electric charges and currents, the fields can be expressed in a similar manner via another pair of potentials, i.e., electric vector potential $\mathbf{C}$ and magnetic scalar potential $\psi$ [6]:

$$\mathbf{E} = -\nabla \times \mathbf{C}, \tag{2.7}$$

$$\mathbf{B} = -\nabla \psi - \frac{\partial \mathbf{C}}{\partial t}. \tag{2.8}$$

Let us now assume that we have neither magnetic nor electric charges and currents. Then the fields can be written both in the form of Eqs. (2.5), (2.6) and in the form of Eqs. (2.7), (2.8). Equating different expressions for $\mathbf{E}$ and $\mathbf{B}$, we obtain the system of equations

$$\frac{\partial \mathbf{A}}{\partial t} - \nabla \times \mathbf{C} + \nabla \varphi = 0, \tag{2.9}$$

$$\frac{\partial \mathbf{C}}{\partial t} + \nabla \times \mathbf{A} + \nabla \psi = 0. \tag{2.10}$$

To close this system, we have to add gauge conditions for the potentials to Eqs. (2.9), (2.10). In particular, choosing the Lorentz gauge, we obtain

$$\frac{\partial \varphi}{\partial t} + \nabla \cdot \mathbf{A} = 0, \tag{2.11}$$

$$\frac{\partial \psi}{\partial t} + \nabla \cdot \mathbf{C} = 0. \tag{2.12}$$

Equations (2.9) – (2.12) give the new formulation of equations of electrodynamics. In contrast to the standard formulation, which involves the vector magnetic potential and the scalar electric potential and yields wave equations for them, the resultant system includes only first-order equations; therefore, initial data for the derivatives of these potentials with respect to time are not needed in formulating the initial-boundary value problem for this system. Moreover, all equations included into system (2.9) – (2.12) are evolutionary equations, and the system has no equations of the form of Eqs. (2.3) – (2.4), which are differential constraints and make the numerical solution of the original system of the Maxwell equations substantially more difficult.

Owing to their construction, the derived equations are valid only if there are no sources (charges and currents) for the field. They allow us to find the free electromagnetic field, which is the solution of the homogeneous Maxwell equations with zero right-hand sides. In the general case, where both electric and magnetic charges and currents are present, in accordance with the superposition principle, the fields are given by the sums of Eqs. (2.6), (2.7) and (2.5), (2.8) [6]:

$$\mathbf{E} = -\nabla \varphi - \frac{\partial \mathbf{A}}{\partial t} - \nabla \times \mathbf{C},$$

$$\mathbf{B} = \nabla \times \mathbf{A} - \nabla \psi - \frac{\partial \mathbf{C}}{\partial t}.$$

Nevertheless, we can formally retain the above-proposed construction in the presence of sources as well. For this purpose, we determine the vector of "vacuum" polarization under the action of electric charges $\mathbf{P}_e$, such that

$$\nabla \cdot \mathbf{P}_e = -\rho_e. \tag{2.13}$$

Then the vector field $\mathbf{E} + \mathbf{P}_e$ is solenoidal, and we can write

$$\mathbf{E} = -\nabla \times \mathbf{C} - \mathbf{P}_e.$$

Let us now introduce the vector of "vacuum" magnetization induced by electric currents, $\mathbf{M}_e$, in accordance with the formula

$$\nabla \times \mathbf{M}_e = \mathbf{J}_e - \frac{\partial \mathbf{P}_e}{\partial t}. \tag{2.14}$$

The fact that the right-hand side of Eq. (2.14) is really a solenoidal vector field follows from the law of conservation of the electric charge. Then Eq. (2.1) yields

$$\mathbf{B} = -\nabla \psi - \frac{\partial \mathbf{C}}{\partial t} + \mathbf{M}_e.$$

Similarly, introducing auxiliary vectors $\mathbf{M}_m$ and $\mathbf{P}_m$, such that

$$\nabla \cdot \mathbf{P}_m = -\rho_m, \tag{2.15}$$

$$-\nabla \times \mathbf{M}_m = \mathbf{J}_m - \frac{\partial \mathbf{P}_m}{\partial t}, \tag{2.16}$$

we obtain expressions for the electric and magnetic fields via the "usual" potentials:

$$\mathbf{E} = -\nabla \varphi - \frac{\partial \mathbf{A}}{\partial t} + \mathbf{M}_m,$$

$$\mathbf{B} = \nabla \times \mathbf{A} - \mathbf{P}_m.$$

Equating, as previously, different expressions for $\mathbf{E}$ and $\mathbf{B}$, we obtain the following relations instead of Eqs. (2.9), (2.10):

$$\frac{\partial \mathbf{A}}{\partial t} - \nabla \times \mathbf{C} + \nabla \varphi = \mathbf{P}_e + \mathbf{M}_m, \tag{2.17}$$

$$\frac{\partial \mathbf{C}}{\partial t} + \nabla \times \mathbf{A} + \nabla \psi = \mathbf{P}_m + \mathbf{M}_e. \tag{2.18}$$

It is obvious, however, that, to determine the vectors $\mathbf{P}_e$, $\mathbf{P}_m$, $\mathbf{M}_e$, and $\mathbf{M}_m$ for specified charges and currents, we have first to solve system (2.13) – (2.16), which actually coincides with the original Maxwell equations. Thus, the generalization of the new approach to the case with non-zero charges and currents is essentially formal. Nevertheless, Eqs. (2.17), (2.18) cannot be considered as completely useless. The point is that electromagnetic field sources are often prescribed in the form of known polarizations and magnetizations. This means that a particular solution of inhomogeneous Maxwell equations is known, which does not necessarily satisfy, however, the imposed initial and boundary conditions. In such a case, the solution of the initial-boundary value problem of interest for us is a superposition of this particular solution and a certain solution of homogeneous equations. This sought solution can be found from system (2.17), (2.18).

We can easily see that the above-derived system of first-order equations for the potentials is invariant with respect to the Lorentz transformations. To write this system in an explicitly relativistic invariant form, we introduce the standard four-dimensional notations

$$x^\mu = \left(x^0, x^1, x^2, x^3\right) = (t, \mathbf{x}), \quad x_\mu = \eta_{\mu\nu} x^\nu = (t, -\mathbf{x})$$

$$\partial_\mu = \frac{\partial}{\partial x^\mu} = \left(\frac{\partial}{\partial t}, \nabla\right), \qquad \partial^\mu = \frac{\partial}{\partial x_\mu} = \left(\frac{\partial}{\partial t}, -\nabla\right).$$

As usually, the covariant and contravariant components of the 4-vectors are related by means of the metric tensor of the Minkowski space

$$\eta_{\mu\nu} = \eta^{\mu\nu} = \text{diag}(1,-1,-1,-1).$$

We also introduce the 4-potentials

$$A^{\mu} = (\varphi, \mathbf{A}), \qquad C^{\mu} = (\psi, \mathbf{C}),$$

the antisymmetric tensor of the electromagnetic field $F^{\mu\nu}$ and the dual tensor $\tilde{F}^{\mu\nu}$:

$$F^{\mu\nu} = \begin{pmatrix} 0 & -E_x & -E_y & -E_z \\ E_x & 0 & -B_z & B_y \\ E_y & B_z & 0 & -B_x \\ E_z & -B_y & B_x & 0 \end{pmatrix}, \qquad \tilde{F}^{\mu\nu} = \frac{1}{2}\varepsilon^{\mu\nu\sigma\tau}F_{\sigma\tau} = \begin{pmatrix} 0 & -B_x & -B_y & -B_z \\ B_x & 0 & E_z & -E_y \\ B_y & -E_z & 0 & E_x \\ B_z & E_y & -E_x & 0 \end{pmatrix}.$$

Here $\varepsilon^{\mu\nu\sigma\tau}$ is an absolutely antisymmetric tensor of the fourth rank (Levi-Civita symbol) that changes its sign in the case of permutation of two arbitrary subscripts and such that $\varepsilon^{0123} = +1$.

The Maxwell equations can be written as a pair of equations

$$\partial_{\mu}F^{\mu\nu} = J_e^{\nu}, \qquad \partial_{\mu}\tilde{F}^{\mu\nu} = J_m^{\nu},$$

where the electric and magnetic 4-currents are

$$J_e^{\nu} = (\rho_e, \mathbf{J}_e), \qquad J_m^{\nu} = (\rho_m, \mathbf{J}_m).$$

We introduce two antisymmetric tensors of the electric $M_e^{\mu\nu}$ and magnetic $M_m^{\mu\nu}$ polarizations of vacuum, which are composed of the vectors $-\mathbf{P}_e, \mathbf{M}_e$ and $-\mathbf{P}_m, -\mathbf{M}_m$, respectively, in the same manner as the electromagnetic field tensor is composed from the vectors $\mathbf{E}$ and $\mathbf{B}$. Then we obtain

$$J_e^{\nu} = \partial_{\mu}M_e^{\mu\nu}, \qquad J_m^{\nu} = \partial_{\mu}M_m^{\mu\nu}.$$

Divergences of the tensor fields $F^{\mu\nu} - M_e^{\mu\nu}$ and $\tilde{F}^{\mu\nu} - M_m^{\mu\nu}$ are equal to zero; therefore, the fields can be expressed via the 4-vectors of two potentials, which yields the formulas

$$F^{\mu\nu} = \partial^{\mu}A^{\nu} - \partial^{\nu}A^{\mu} + M_e^{\mu\nu}, \qquad \tilde{F}^{\mu\nu} = \partial^{\mu}C^{\nu} - \partial^{\nu}C^{\mu} + M_m^{\mu\nu}.$$

Using the duality of the tensors $F^{\mu\nu}$ and $\tilde{F}^{\mu\nu}$, one can write the relations

$$\partial^{\mu}C^{\nu} - \partial^{\nu}C^{\mu} + N_m^{\mu\nu} = \frac{1}{2}\varepsilon^{\mu\nu\sigma\tau}\left(\partial_{\sigma}A_{\tau} - \partial_{\tau}A_{\sigma} + M_{e,\sigma\tau}\right), \tag{2.19}$$

$$\partial^{\mu}A^{\nu} - \partial^{\nu}A^{\mu} + N_e^{\mu\nu} = \frac{1}{2}\varepsilon^{\mu\nu\sigma\tau}\left(\partial_{\sigma}C_{\tau} - \partial_{\tau}C_{\sigma} + M_{m,\sigma\tau}\right). \tag{2.20}$$

Relations (2.19) and (2.20) are equivalent, both being four-dimensional forms of Eqs. (2.17) and (2.18). These relations can be supplemented by equations that express the Lorentz gauge condition for the potentials:

$$\partial_{\mu}A^{\mu} = 0, \qquad \partial_{\mu}C^{\mu} = 0.$$

## 3. TWO-POTENTIAL FORMALISM FOR THE ELECTROMAGNETIC FIELD IN THE MEDIUM

It is known that the Maxwell equations (2.1-2.4) remains valid in a continuous medium as well if the charges and currents in the right-hand sides of the equations are understood as sums of free and bound charges and currents. It is more convenient, however, to retain only the densities of free charges and currents in the right-hand sides. For this purpose, the bound charges and currents are expressed via the medium polarization and magnetization vectors $\bar{\mathbf{P}}_e$, $\bar{\mathbf{P}}_m$, $\bar{\mathbf{M}}_e$, $\bar{\mathbf{M}}_m$, as it was done in Section 2 for free charges and currents in vacuum. Introducing auxiliary vectors

$$\mathbf{D} = \mathbf{E} + \bar{\mathbf{P}}_e, \qquad \bar{\mathbf{E}} = \mathbf{E} - \bar{\mathbf{M}}_m$$
$$\mathbf{H} = \mathbf{B} - \bar{\mathbf{M}}_e, \qquad \bar{\mathbf{B}} = \mathbf{B} + \bar{\mathbf{P}}_m.$$

we write the Maxwell equations in the medium as

$$\frac{\partial \mathbf{D}}{\partial t} - \nabla \times \mathbf{H} = -\mathbf{J}_e, \tag{3.1}$$

$$\frac{\partial \bar{\mathbf{B}}}{\partial t} + \nabla \times \bar{\mathbf{E}} = -\mathbf{J}_m, \tag{3.2}$$

$$\nabla \cdot \mathbf{D} = \rho_e, \tag{3.3}$$

$$\nabla \cdot \bar{\mathbf{B}} = \rho_m, \tag{3.4}$$

where the densities of charges and currents in the right-hand sides are understood as the densities of free charges and currents only. In what follows, the bar above the vectors $\bar{\mathbf{E}}$ and $\bar{\mathbf{B}}$ is omitted because their difference from $\mathbf{E}$ and $\mathbf{B}$ is substantial only in hypothetical media containing magnetic monopoles.

In contrast to the Maxwell equations in vacuum, introduction of polarization and magnetization vectors and transfer of the bound source fields from the right-hand side to the left-hand side of the equations are not purely formal operations. The point is that the vectors $\mathbf{D}$ and $\mathbf{H}$ in most of various media can be related to the vectors $\mathbf{E}$ and $\mathbf{B}$ by fairly simple expressions. In this paper, we confine ourselves to considering linear isotropic media without dispersion, which are not necessarily homogeneous. For such media, the relations take a particularly simple form

$$\mathbf{D} = \varepsilon \mathbf{E}, \qquad \mathbf{B} = \mu \mathbf{H}, \tag{3.5}$$

though the approach considered here can be rather easily extended to more complicated (in particular, anisotropic and nonlinear) media. In Eqs. (3.5), the relative permittivity $\varepsilon$ and the the relative permeability $\mu$ of the material are known functions of spatial coordinates; in this form, these relations describe many media of practical importance, in particular, photon crystals.

The vector and scalar electric potentials are introduced with the help of the same definitions as those used in Section 2; the magnetic potentials are described by the formulas

$$\mathbf{D} = -\nabla \times \mathbf{C} - \bar{\mathbf{P}}_e,$$
$$\mathbf{H} = -\nabla \psi - \frac{\partial \mathbf{C}}{\partial t} + \bar{\mathbf{M}}_e.$$

Substituting the expressions of the field via the electric and magnetic potentials into the material relations (3.5), we obtain the system

$$\frac{\partial \mathbf{A}}{\partial t} - \varepsilon^{-1}\nabla \times \mathbf{C} + \nabla \varphi = \varepsilon^{-1}\overline{\mathbf{P}}_e + \overline{\mathbf{M}}_m, \qquad (3.6)$$

$$\frac{\partial \mathbf{C}}{\partial t} + \mu^{-1}\nabla \times \mathbf{A} + \nabla \psi = \mu^{-1}\overline{\mathbf{P}}_m + \overline{\mathbf{M}}_e. \qquad (3.7)$$

Thus, Eqs. (3.6) and (3.7) can be considered as extension of Eqs. (2.17) and (2.18) to the case of electromagnetic processes in a material medium. The polarization and magnetization vectors in the right-hand sides of these equations are related only to free charges and currents.

To close the system of equations, we have to supplement Eqs. (3.6) and (3.7) with gauge conditions for the potentials. They are usually chosen in such a manner that the resultant second-order equations for the potentials are as simple as possible. The natural choice for a homogeneous medium is [6]

$$\varepsilon\mu \frac{\partial \varphi}{\partial t} + \nabla \cdot \mathbf{A} = 0, \qquad (3.8)$$

$$\varepsilon\mu \frac{\partial \psi}{\partial t} + \nabla \cdot \mathbf{C} = 0, \qquad (3.9)$$

which yields simple wave equations for the potentials. In the general case of a medium with $\varepsilon$ and $\mu$ dependent on spatial variables, the choice of the "best" gauge is not obvious. This problem was discussed in [7-9]. The most consistent approach to obtaining the gauge conditions in a material medium seems to be writing the material relations (3.5) in a relativistically invariant form (for this purpose, it is necessary to introduce a fourth rank tensor dependent on the permittivity and the permeability and 4-velocity of the medium in order to describe the medium properties) and deriving an equation for the 4-potential in a moving medium. The corresponding calculations can be found in [8]. The resultant equation is appreciably simplified if the gauge condition is chosen in the form

$$\partial_\mu A^\mu + (\varepsilon\mu - 1)U^\mu U_\nu \partial_\mu A^\nu = 0, \qquad U^\mu = \left(\frac{1}{\sqrt{1-\mathbf{u}^2}}, \frac{\mathbf{u}}{\sqrt{1-\mathbf{u}^2}}\right).$$

Here $U^\mu$ is the 4-velocity vector and $\mathbf{u}$ is the three-dimensional velocity of the medium. For a quiescent medium, this condition reduces to Eq. (3.8). Based on this fact, we take Eqs. (3.8) and (3.9) as gauge conditions for the potentials. Together with Eqs. (3.6) and (3.7), they form a closed system of equations for the electromagnetic potentials in the medium.

## 4. SOME MATHEMATICAL PROPERTIES OF THE DERIVED EQUATIONS

Introducing the "quasi-vector"

$$\mathbf{U} = \left(D_x, D_y, D_z, B_x, B_y, B_z\right)^T$$

and using the linear material relations (3.5), the Maxwell equations (3.1), (3.2) without the right-hand sides can be rewritten in the matrix form as

$$\frac{\partial \mathbf{U}}{\partial t} + \mathbf{A}_x \frac{\partial \mathbf{U}}{\partial x} + \mathbf{A}_y \frac{\partial \mathbf{U}}{\partial y} + \mathbf{A}_z \frac{\partial \mathbf{U}}{\partial z} = 0. \qquad (4.1)$$

The most important property of equations involving the curls of the electric and magnetic fields is the fact that they form a hyperbolic system. By definition, system (4.1) is hyperbolic [10] if, for an arbitrary unit vector $\mathbf{n} = (n_x, n_y, n_z)$, the eigenvalues of the matrix

$$\mathbf{A} = n_x \mathbf{A}_x + n_y \mathbf{A}_y + n_z \mathbf{A}_z = \begin{pmatrix} 0 & 0 & 0 & 0 & n_z/\mu & -n_y/\mu \\ 0 & 0 & 0 & -n_z/\mu & 0 & n_x/\mu \\ 0 & 0 & 0 & n_y/\mu & -n_x/\mu & 0 \\ 0 & -n_z/\varepsilon & n_y/\varepsilon & 0 & 0 & 0 \\ n_z/\varepsilon & 0 & -n_x/\varepsilon & 0 & 0 & 0 \\ -n_y/\varepsilon & n_x/\varepsilon & 0 & 0 & 0 & 0 \end{pmatrix}$$

are real and it can be brought to a diagonal form by applying a similarity transformation (i.e., there exists a complete system of eigenvectors). The eigenvalues of $\mathbf{A}$ are [11]

$$\lambda_{1,2} = -c, \quad \lambda_{3,4} = 0, \quad \lambda_{5,6} = c, \quad \text{where} \quad c = 1/\sqrt{\varepsilon\mu}.$$

As a rule, it is the system (3.1), (3.2) that is integrated with respect to time in solving the Maxwell equations numerically, while Eqs. (3.3) and (3.4) are not considered at all. The latter equations are differential constraints, actually restricting the class of admissible initial data. If the electric and magnetic fields are solenoidal at the initial time moment, then these fields remain solenoidal at all subsequent time moment in the case of the *exact* solution of the differential equations (3.1) and (3.2). In the numerical solution, however, this property can be violated, and non-physical electric and magnetic charges can appear inside the computational domain. This is one of the most severe problems in numerical simulation of electromagnetic processes. Various techniques were proposed to overcome this problem; in particular, they include the correction of the fields by solving Poisson's equation at each time step [12], the use of modified (so-called perfectly hyperbolic) Maxwell equations [13], the use of extended overdetermined hyperbolic systems for correcting the solution [14], etc.

The above-mentioned problem is directly related to the mathematical properties of system (3.1), (3.2), namely, to the existence of a zero eigenvalue (of multiplicity equal to 2). Indeed, the modes corresponding to this eigenvalue are non-physical: electromagnetic waves always propagate in nature with the speed of light, which corresponds to the eigenvalues $\pm c$. It can be easily demonstrated that the subspace corresponding to the zero eigenvalue consists of vector fields with the curl equal to zero and, thus, with the divergence that differs from zero.

Let us now consider Eqs. (3.6)-(3.9), which describe the electromagnetic fields in the new two-potential formalism. They are written in the matrix form as

$$\frac{\partial \mathbf{U}}{\partial t} + \mathbf{A}_x \frac{\partial \mathbf{U}}{\partial x} + \mathbf{A}_y \frac{\partial \mathbf{U}}{\partial y} + \mathbf{A}_z \frac{\partial \mathbf{U}}{\partial z} = 0, \tag{4.2}$$

where

$$\mathbf{U} = (A_x, A_y, A_z, \psi, C_x, C_y, C_z, \varphi)^T.$$

In this case, the matrix $\mathbf{A} = n_x \mathbf{A}_x + n_y \mathbf{A}_y + n_z \mathbf{A}_z$ is written as

$$\mathbf{A} = \begin{pmatrix} 0 & 0 & 0 & 0 & 0 & n_z/\varepsilon & -n_y/\varepsilon & n_x \\ 0 & 0 & 0 & 0 & -n_z/\varepsilon & 0 & n_x/\varepsilon & n_y \\ 0 & 0 & 0 & 0 & n_y/\varepsilon & -n_x/\varepsilon & 0 & n_z \\ 0 & 0 & 0 & 0 & n_x/\varepsilon\mu & n_y/\varepsilon\mu & n_z/\varepsilon\mu & 0 \\ 0 & -n_z/\mu & n_y/\mu & n_x & 0 & 0 & 0 & 0 \\ n_z/\mu & 0 & -n_x/\mu & n_y & 0 & 0 & 0 & 0 \\ -n_y/\mu & n_x/\mu & 0 & n_z & 0 & 0 & 0 & 0 \\ n_x/\varepsilon\mu & n_y/\varepsilon\mu & n_z/\varepsilon\mu & 0 & 0 & 0 & 0 & 0 \end{pmatrix}.$$

After calculations, we find that the matrix $\mathbf{A} = n_x \mathbf{A}_x + n_y \mathbf{A}_y + n_z \mathbf{A}_z$ has two eigenvalues, each having multiplicity equal to 4:

$$\lambda_{1,2,3,4} = -c, \qquad \lambda_{5,6,7,8} = c.$$

The left-hand side of Eqs. (4.2) can be transformed to the following characteristic form (that determines the choice of the left eigenvectors, which is not unique, generally speaking):

$$\lambda = -c: \quad \begin{cases} \left( \mathbf{n} \dfrac{\partial \psi}{\partial t} - c \dfrac{\partial \mathbf{C}}{\partial t} + \dfrac{1}{\mu} \mathbf{n} \times \dfrac{\partial \mathbf{A}}{\partial t} \right) - c \left( \mathbf{n} \dfrac{\partial \psi}{\partial \xi} - c \dfrac{\partial \mathbf{C}}{\partial \xi} + \dfrac{1}{\mu} \mathbf{n} \times \dfrac{\partial \mathbf{A}}{\partial \xi} \right) = 0, \\[6pt] \left( \dfrac{\partial \varphi}{\partial t} - c \mathbf{n} \cdot \dfrac{\partial \mathbf{A}}{\partial t} \right) - c \left( \dfrac{\partial \varphi}{\partial \xi} - c \mathbf{n} \cdot \dfrac{\partial \mathbf{A}}{\partial \xi} \right) = 0, \end{cases}$$

$$\lambda = +c: \quad \begin{cases} \left( \mathbf{n} \dfrac{\partial \varphi}{\partial t} + c \dfrac{\partial \mathbf{A}}{\partial t} - \dfrac{1}{\varepsilon} \mathbf{n} \times \dfrac{\partial \mathbf{C}}{\partial t} \right) + c \left( \mathbf{n} \dfrac{\partial \varphi}{\partial \xi} + c \dfrac{\partial \mathbf{A}}{\partial \xi} - \dfrac{1}{\varepsilon} \mathbf{n} \times \dfrac{\partial \mathbf{C}}{\partial \xi} \right) = 0, \\[6pt] \left( \dfrac{\partial \psi}{\partial t} + c \mathbf{n} \cdot \dfrac{\partial \mathbf{C}}{\partial t} \right) + c \left( \dfrac{\partial \psi}{\partial \xi} + c \mathbf{n} \cdot \dfrac{\partial \mathbf{C}}{\partial \xi} \right) = 0, \end{cases}$$

Here $\xi$ is the coordinate along the direction determined by the vector $\mathbf{n}$. Thus, in contrast to system (4.1), the equations written with the use of the two-potential formalism do not involve the zero eigenvalue; all modes propagate with the speed of light, as it should be. The Lorentz gauge conditions (3.8) and (3.9) for the potentials play here an important role. Indeed, the equations for the potentials can be appreciably simplified if the Coulomb gauge is chosen:

$$\varphi = \psi = 0, \qquad \nabla \cdot \mathbf{A} = 0, \qquad \nabla \cdot \mathbf{C} = 0.$$

In this case, however, the remaining six equations for two vector potentials have exactly the same structure as the Maxwell equations with the curls (3.1) and (3.2) with all above-mentioned problems of maintenance of the fields $\mathbf{A}, \mathbf{C}$ solenoidal and non-physical modes propagating with zero velocity.

## 5. EXAMPLES OF USING THE FORMALISM OF TWO POTENTIALS IN NUMERICAL SIMULATIONS

The numerical solution of Eqs. (3.6)-(3.9) can be performed with any method among the vast pool of techniques that have been developed for hyperbolic systems of equations [15]. In this work, the spatial approximation is performed by one of the modern shock-capturing methods: the so-called fifth-order Weighted Essentially Non-Oscillatory (WENO) scheme [16]. This scheme, which can be considered as a remote descendant of the famous Godunov scheme [17], has been well approved in solving problems of supersonic aerodynamics (see, e.g., [18]). We will not discuss here all specific features of using the WENO scheme for the numerical solution of the above-described equations of the new formalism, because we hope to describe them in a separate paper.

Integration with respect to time was performed by the fourth-order Runge-Kutta-Gill method [19], which requires a smaller amount of computer memory for storage of auxiliary arrays than the "standard" fourth-order Runge-Kutta method.

As the first example, we consider propagation of an electromagnetic wave in a metallic waveguide with a square cross section. The initial data are defined in the form of a transversely electric $TE_{1,1}$ mode of the waveguide [20] with a frequency $\omega = \pi\sqrt{6}$.

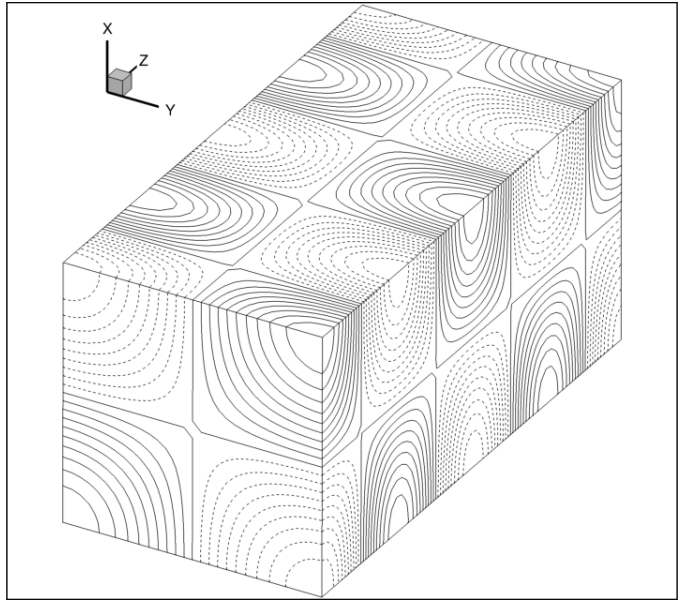

**Fig. 1.**

The problem is solved in a computational domain $0 \le x \le 1$, $0 \le y \le 1$, $0 \le z \le 2$ on a grid of 30×30×60 cells. Symmetry or antisymmetry conditions are imposed on various components of the potentials on the waveguide surface to guarantee satisfaction of the usual conditions for the components of the electric and magnetic fields on an infinitely conducting surface. Periodic boundary conditions are applied in the longitudinal direction at $z = 0$ and 1. The step of integration with respect to time is determined from the condition $\Delta t = CFL(1/\Delta x + 1/\Delta y + 1/\Delta z)$, where the Courant-Friedrichs-Lewy number is $CFL = 0.8$.

Figure 1 shows the computed distribution of the longitudinal magnetic field $B_z$ on the waveguide surface and in its end cross section at the time $t = 2\pi/\omega = 0.8165$. The electric and magnetic fields are calculated from the vector potentials in accordance with Eqs. (2.5) and (2.7); a sixth-order central-difference scheme is used for differentiation.

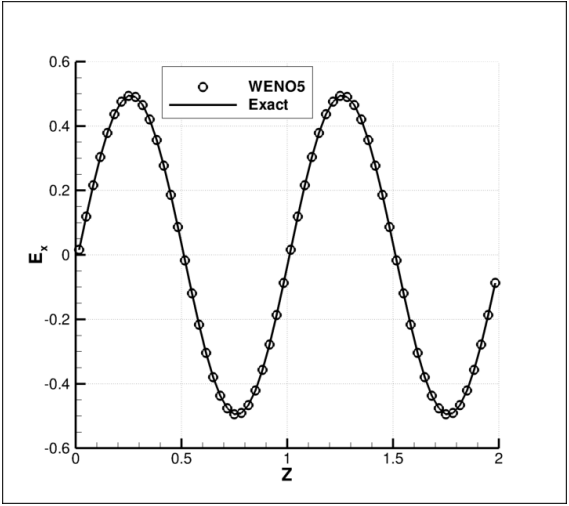 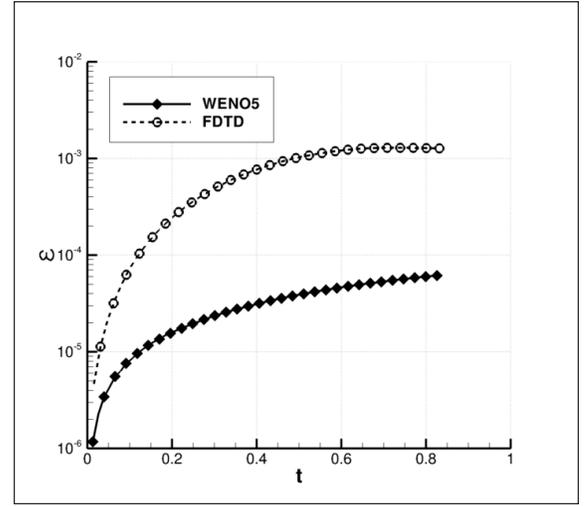

**Fig. 2.**  **Fig. 3.**

The computed distribution of the electric field component $E_x$ along the line $x = y = 1/3$ at the same time moment is compared in Fig. 2 with the exact solution. It is seen that these solutions cannot be distinguished in the scale of this figure.

To get a better idea about the accuracy ensured by the WENO scheme, we also solved the same problem using the Finite-Difference Time Domain (FDTD) method, which is most popular in computational electrodynamics. This method is a second-order finite-difference scheme on a staggered space-time grid [1]. Figure 3 shows the time evolution of the error (in the norm $L_1$) of determining $B_z$ for the two compared methods. As is seen from Fig. 3, the WENO scheme is more accurate than the FDTD method by an order of magnitude.

In the second example, we consider propagation of an electromagnetic wave in a medium with a jump of material properties. We model the normal incidence (in the $x$ direction) of a plane-polarized wave onto the interface between two media with different values of dielectric permittivity. The problem is solved in the domain $0 \leq z \leq 2$. We set $\varepsilon = 4$, and $\varepsilon = 1$ on the left and on the right of the plane $z = 1$. The magnetic permeability is $\mu = 1$ everywhere. The domain is divided into 400 cells. At the initial time, the field is equal to zero everywhere inside the computational domain. On the left boundary, we impose harmonically varying (with a frequency $\omega = 10$) values of the potentials corresponding to a plane electromagnetic wave traveling to the right. The calculation is performed with $CFL = 0.8$ up to the time $t = 3$ when the wave reaches the right boundary of the computational domain. At this time, the solution in the interval $0 < z < 0.5$ includes only the incident wave, the solution in the interval $0.5 < z < 1$ is a superposition of the incident and reflected waves, and the solution at $z > 1$ consists of the wave that passed to the second medium.

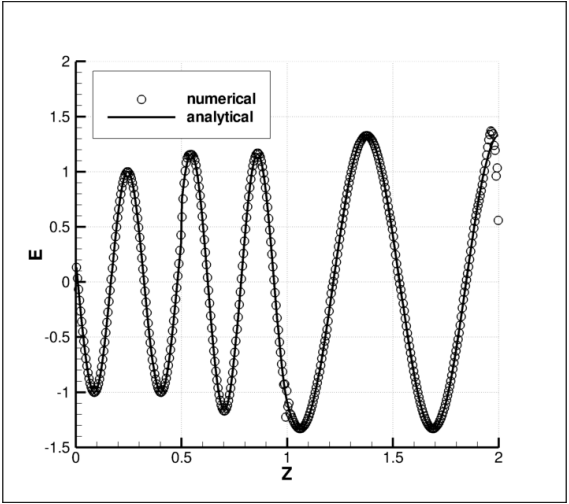 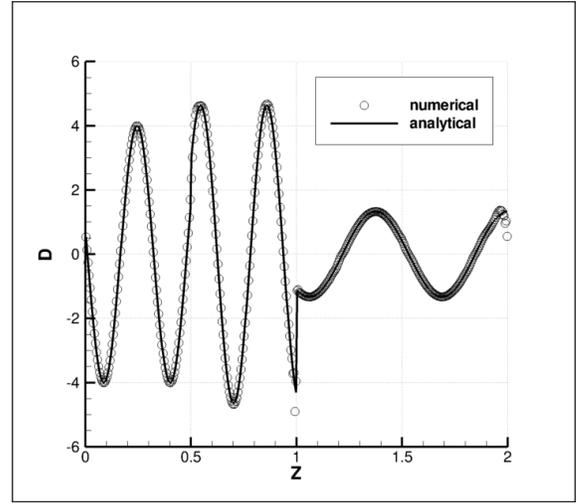

**Fig. 4.**   **Fig. 5.**

The distributions of the only non-zero component of the electric field intensity $E_x$ and the corresponding components of the vector of electric induction $D_x$ at the final time instant are compared in Figs. 4 and 5 with exact analytical expressions for them.

Obviously, the numerical and analytical solutions are in good agreement. As it could be expected, the tangential component of the electric field intensity is continuous on the interface between two media (Fig. 4), whereas the tangential component of the electric induction vector has a jump (Fig. 5). Small differences between the numerical and analytical solutions are observed only on the interface and near the leading front of the transmitted wave around $z = 2$. These differences are caused by using a central-difference formula for calculating the fields from the potentials. At the above-indicated points, the seven-point stencil of this formula intersects grid points where the differentiated function has a kink, which leads to a decrease in accuracy. Obviously, the results can be improved by using one-sided difference formulas that do not intersect the kink points of the differentiated function for calculating the fields at such points.

These considerations are illustrated in Figs. 6 and 7, which show the distributions of the component $C_y$ of the electric vector potential and the component $A_x$ of the magnetic vector potential, respectively, at $t = 3$. The knik in the distribution of $C_y$ on the interface of two media, which corresponds to a jump of the tangential component $D_x$ of the electric induction vector, is clearly visible. At the same time, the distribution of $A_x$ looks smooth, as it should be because there is no jump of the tangential component of the magnetic field $B_y$ on the interface of two media with an identical value of $\mu$. Figures 6 and 7 illustrate one more advantage of formulating the electromagnetism equations with the use of potentials: they are smoother than the fields themselves. It should be noted that the computation of propagation of electromagnetic waves in a medium with discontinuities of dielectric permittivity by FDTD-type methods based on approximation of spatial derivatives by central differences without involving special techniques (e.g., replacement of the jump of material properties by a zone of their rapid, but still continuous variation) will inevitably lead to emergence of noticeable non-physical oscillations of the numerical solution.

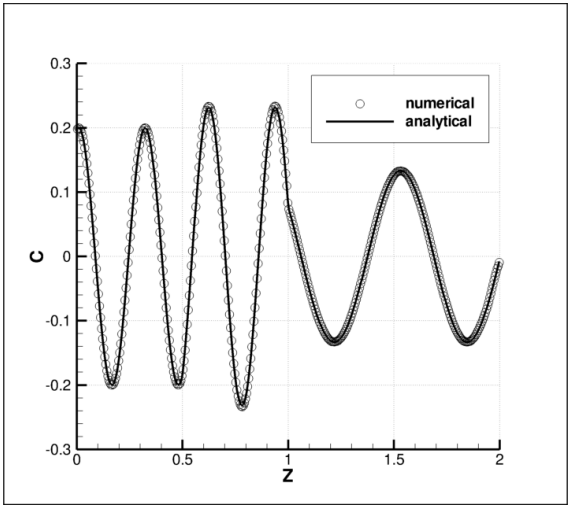 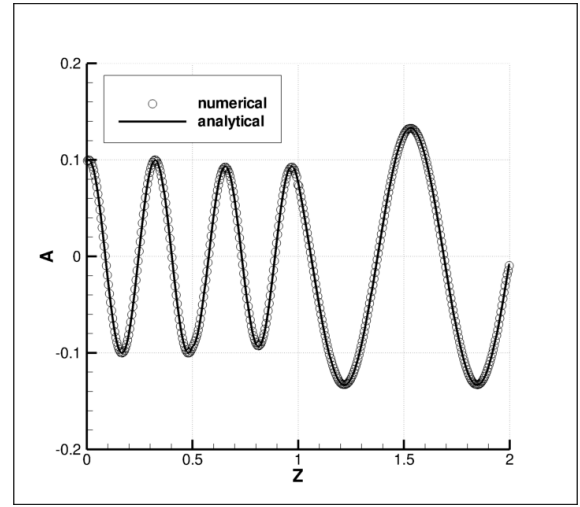

**Fig. 6.**  **Fig. 7.**

## 6. CONCLUSIONS

Thus, a new formulation of electromagnetic field equations based on the use of two vector potentials and two scalar potentials is proposed. This formulation allows the Maxwell equations both in vacuum and in a material medium to be written in the form of a hyperbolic system possessing a number of desirable properties. In particular, it consists only of evolutionary equations and has no relations having the character of differential constraints and leading to significant problems in the numerical solution of the Maxwell equations in the standard formulation. Moreover, all eigenvalues of the Jacobi matrix of the derived system of equations corresponds to physical modes propagating with the speed of light; there are no non-physical modes corresponding to the zero eigenvalue and obtained in the frequently used approach where only the equations containing curls of the vector fields are solved, whereas the equations with divergences are ignored. All these facts allow powerful modern shock-capturing methods based on approximation of spatial derivatives by upwind differences to be used to solve the new system numerically.

Examples of numerical simulations of propagation of electromagnetic waves by solving the equations in the new formulation are given. One of the modern schemes is used: a fifth-order WENO scheme. It is demonstrated that such a numerical approach allows the solution to be obtained with high accuracy, including problems that involve jumps of material properties of the medium.